\documentclass[conference]{IEEEtran}
\usepackage{graphicx}
\ifCLASSINFOpdf
\else
\fi

\usepackage{caption}
\usepackage{url}
\usepackage{amsmath}
\usepackage{hyperref}
\usepackage{amssymb}
\usepackage{listings}
\usepackage{xcolor}
\definecolor{codeblue}{rgb}{0.1, 0.1, 0.8}
\definecolor{codegray}{rgb}{0.5, 0.5, 0.5}
\definecolor{codepurple}{rgb}{0.58, 0, 0.82}
\definecolor{backcolour}{rgb}{0.95, 0.95, 0.92}

\lstdefinestyle{mystyle}{
    backgroundcolor=\color{backcolour},   
    commentstyle=\color{codegray},
    keywordstyle=\color{codeblue},
    numberstyle=\tiny\color{codegray},
    stringstyle=\color{codepurple},
    basicstyle=\ttfamily\footnotesize,
    breakatwhitespace=false,         
    breaklines=true,                 
    captionpos=b,                    
    keepspaces=true,                 
    numbers=left,                    
    numbersep=5pt,                  
    showspaces=false,                
    showstringspaces=false,
    showtabs=false,                  
    tabsize=2
}

\begin{document}
%
\title{oneDAL Optimization for ARM Scalable Vector  \\
Extension: Maximizing Efficiency for High-Performance Data Science}
\author{
    \IEEEauthorblockN{.\\.}
    \IEEEauthorblockA{.\\
                      .\\.}
}

\author{
    \IEEEauthorblockN{Chandan Sharma, Rakshith GB, Ajay Kumar Patel, Dhanus M Lal, \\Darshan Patel, Ragesh Hajela, Masahiro Doteguchi, Priyanka Sharma}
    \IEEEauthorblockA{Fujitsu Research of India Pvt. Ltd.\\
    Email: \{chandan.sharma, rakshith.gb, ajaykumar.patel, dhanus.mlal, \\darshan.patel, ragesh.hajela, masahiro.doteguchi, priyanka.s\}@fujitsu.com}
}

%


\maketitle

\begin{abstract}
The evolution of ARM-based architectures, particularly those incorporating Scalable Vector Extension (SVE), has introduced transformative opportunities for high-performance computing (HPC) and machine learning (ML) workloads. The Unified Acceleration Foundation’s (UXL’s) oneAPI Data Analytics Library (oneDAL) is a widely adopted library for accelerating ML and data analytics workflows, but its reliance on Intel®’s proprietary Math Kernel Library (MKL) has traditionally limited its compatibility to x86 platforms. This paper details the porting of oneDAL to ARM architectures with SVE support, using OpenBLAS as an alternative backend to overcome architectural and performance challenges.

Beyond porting, the research introduces novel ARM-specific optimizations, including custom sparse matrix routines, vectorized statistical functions, and a Scalable Vector Extension (SVE)-optimized Support Vector Machine (SVM) algorithm. The SVM enhancements leverage SVE’s flexible vector lengths and predicate-driven execution, achieving notable performance gains 22\% for the Boser method and 5\% for the Thunder method. Benchmarks conducted on ARM SVE-enabled AWS Graviton3 instances showcase up to 200X acceleration in ML training and inference tasks compared to the original scikit-learn implementation on ARM platform.

Moreover, the ARM-optimized oneDAL achieves performance parity with, and in some cases exceeds, the x86 oneDAL implementation (MKL backend) on Ice Lake x86 systems, which are nearly twice as costly as AWS Graviton3 ARM instances. These findings highlight ARM’s potential as a high-performance, energy-efficient platform for data-intensive ML applications. By expanding cross-architecture compatibility and contributing to the open-source ecosystem, this work reinforces ARM’s position as a competitive alternative in the HPC and ML domains, paving the way for future advancements in data-intensive computing.

\end{abstract}




%
\IEEEpeerreviewmaketitle

\section{Introduction}
The growing power of computing fuels scientific discovery and technical advancements, enabling enhanced applications in machine learning and big data analytics, with high-performance computing (HPC) systems depending on efficient hardware and software designs to drive these improvements \cite{hpcforbigdata}. X86-based architectures have historically dominated the HPC environment because of their software availability, robust performance, and wide ecosystem support \cite{hpcmarketshare}. However, this landscape is being redefined by the latest developments in ARM-based processors, such as the A64FX and FX700\cite{fugaku1}. Both the FX700 and the A64FX, which are powerful candidates for next-generation HPC systems because of their exceptional performance and energy efficiency, are employed in the Fugaku supercomputer\cite{fugaku2}. The upcoming ARM processors\cite{ARM, armprocessorwiki, monakasc23, Monakahipc}, which foresee advancements in the future and make use of complex microarchitecture to give improved performance and energy efficiency, further this trend. Energy-efficient computing is crucial to meet the increasing demands of data centers, and ARM CPUs represent a significant advancement in this area. This idea is aligns with global environmental objectives, aiming to significantly reduce energy consumption in data centers \cite{uxlmonaka}.

Among the significant contributions to the HPC and ML ecosystem, the oneAPI Data Analytics Library (oneDAL) \cite{onedal}, developed under the UXL \cite{uxl}. oneDAL provides highly optimized routines for data preprocessing, transformation, analysis, modeling, and validation, leveraging hardware acceleration techniques such as parallelism and vectorized computations. Widely integrated into Python-based workflows through daal4py \cite{daal4py}, oneDAL enables seamless adoption within popular data science and machine learning framework such as scikit-learn \cite{scikitlearn}. However, oneDAL’s reliance on MKL \cite{mkl} restricts its applicability to x86 platforms \cite{systemreqonedal}, creating challenges for adoption on ARM systems.

As ARM processors continue to gain prominence in the HPC landscape, adapting oneDAL for ARM platforms is essential for achieving high-performance data analytics across a broader spectrum of hardware. ARM SVE introduces architectural innovations that overcome limitations inherent in traditional SIMD (Single Instruction, Multiple Data) architectures, such as Intel’s AVX \cite{avx} and ARM’s NEON \cite{armneon}. Unlike fixed-width SIMD systems, SVE employs a Vector Length Agnostic (VLA) approach, allowing vector lengths to scale flexibly between 128 and 2048 bits in increments of 128 bits. This adaptability enables a single, vectorized implementation to operate efficiently across different ARM processors without architecture-specific modifications, making SVE particularly suitable for scientific and data analytics applications that demand high parallelism and scalability.


A defining characteristic of Scalable Vector Extension (SVE) \cite{armsve} is that it is a separate extension from Advanced SIMD (commonly known as Neon), with an entirely new set of instruction encodings specifically designed to address the needs of high-performance computing (HPC). Unlike Neon, which is more general-purpose, SVE is tailored to overcome traditional barriers to auto-vectorization in compilers, enabling more efficient and scalable parallel processing. SVE introduces several key features that significantly enhance flexibility and performance in vectorized computations,  making it particularly suited for data-intensive applications such as machine learning and scientific computing, where complex and irregular data patterns are common. These features include:

\begin{itemize}

\item \textbf{Gather-load and scatter-store} instructions for efficient memory access patterns.
\item \textbf{Per-lane predication}, enabling independent operations on different vector elements.
\item \textbf{Predicate-driven loop control and management}, providing fine-grained control over iterations in vectorized loops.
\end{itemize}

To implement its advanced capabilities, SVE introduces two new types of registers designed to enhance flexibility and efficiency in vector processing:

\textbf{Z Registers:} SVE provides 32 Z registers, which are configurable in width and can hold data elements interpreted as 8-bit, 16-bit, 32-bit, or 64-bit values. Each Z register's width can range up to 2048 bits, with a 2048-bit Z register capable of holding 256 8-bit elements, 128 16-bit elements, 64 32-bit elements, or 32 64-bit elements. This adaptability allows Z registers to handle large datasets in parallel, optimizing performance for data-intensive applications, such as ML algorithms, where multiple elements are processed simultaneously.

\textbf{P Registers:} SVE also introduces 15 predicate registers (P registers) that enable fine-grained control over vector operations by acting as masks. P registers serve the following purposes:
\begin{itemize}
    \item \textbf{P0–P7}: Used as governing predicates for load/store and arithmetic operations.
    \item \textbf{P8–P15}: Serve additional roles in loop management and control.
    \item \textbf{FFR (First Fault Register)}: Provides hardware support for speculative execution and assists in fault recovery during vectorized operations.
\end{itemize}

The flexibility provided by SVE’s Vector Length Agnostic (VLA) processing model, combined with the selective computation capabilities enabled by P registers, makes SVE particularly suited for HPC and ML applications.

Adapting oneDAL for ARM’s SVE architecture involves several challenges. First, a new backend needs to replace MKL, as it is incompatible with ARM systems. OpenBLAS \cite{openblas}, an open-source alternative, provides fundamental BLAS and LAPACK functionalities for ARM, but lacks the advanced optimizations that MKL offers on x86 platforms. Thus, specific ARM-compatible implementations and optimizations are necessary to achieve comparable performance. Second, to fully leverage ARM SVE, key components of oneDAL must be re-engineered to exploit SVE’s unique capabilities, including the vectorized ML algorithm, optimized statistical functions, and custom sparse matrix routines. By implementing ARM-specific enhancements oneDAL can achieve efficient computation for ML tasks on ARM platforms.


\section{Related Work}

The oneAPI initiative provides a unified programming model designed to support diverse hardware architectures, including CPUs, GPUs, and FPGAs. As part of this ecosystem, oneDNN has already been successfully ported to ARM, demonstrating substantial performance improvements on ARM-based platforms \cite{armonednn}. This success paved the way for the adaptation of the oneAPI Data Analytics Library (oneDAL) to ARM, allowing it to leverage ARM's advanced architectures for high-performance data analytics while broadening its compatibility beyond x86 and MKL systems.

Original scikit-learn pipelines include libraries like NumPy \cite{numpy} and SciPy \cite{scipy}, which utilize OpenBLAS for numerical computations. \textit{daal4py} acts as a bridge between scikit-learn and oneDAL, enabling accelerated machine learning (ML) algorithms through the scikit-learn-intelex patch \cite{onedalpatch}. However \textit{daal4py} heavily relies on MKL for its performance optimizations, limiting its applicability to x86 platforms.


The oneAPI initiative also includes SYCL, a cross-platform parallel programming model designed to manage heterogeneous hardware. SYCL facilitates portability across GPUs, CPUs, and other accelerators, offering a unified framework for efficient execution \cite{syclwiki}. However, the implementation of SYCL in oneDAL currently requires a compatible compiler, such as Intel’s DPC++ \cite{dcp}, which is outside the scope of this research. This work focuses on extending oneDAL for ARM architectures, specifically using OpenBLAS and ARM SVE. Despite this, researchers continues to monitor developments in SYCL and DPC++ within the Unified Acceleration Foundation (UXL) to explore potential future integration.

\begin{figure}[h] \centering \includegraphics[width=1\linewidth]{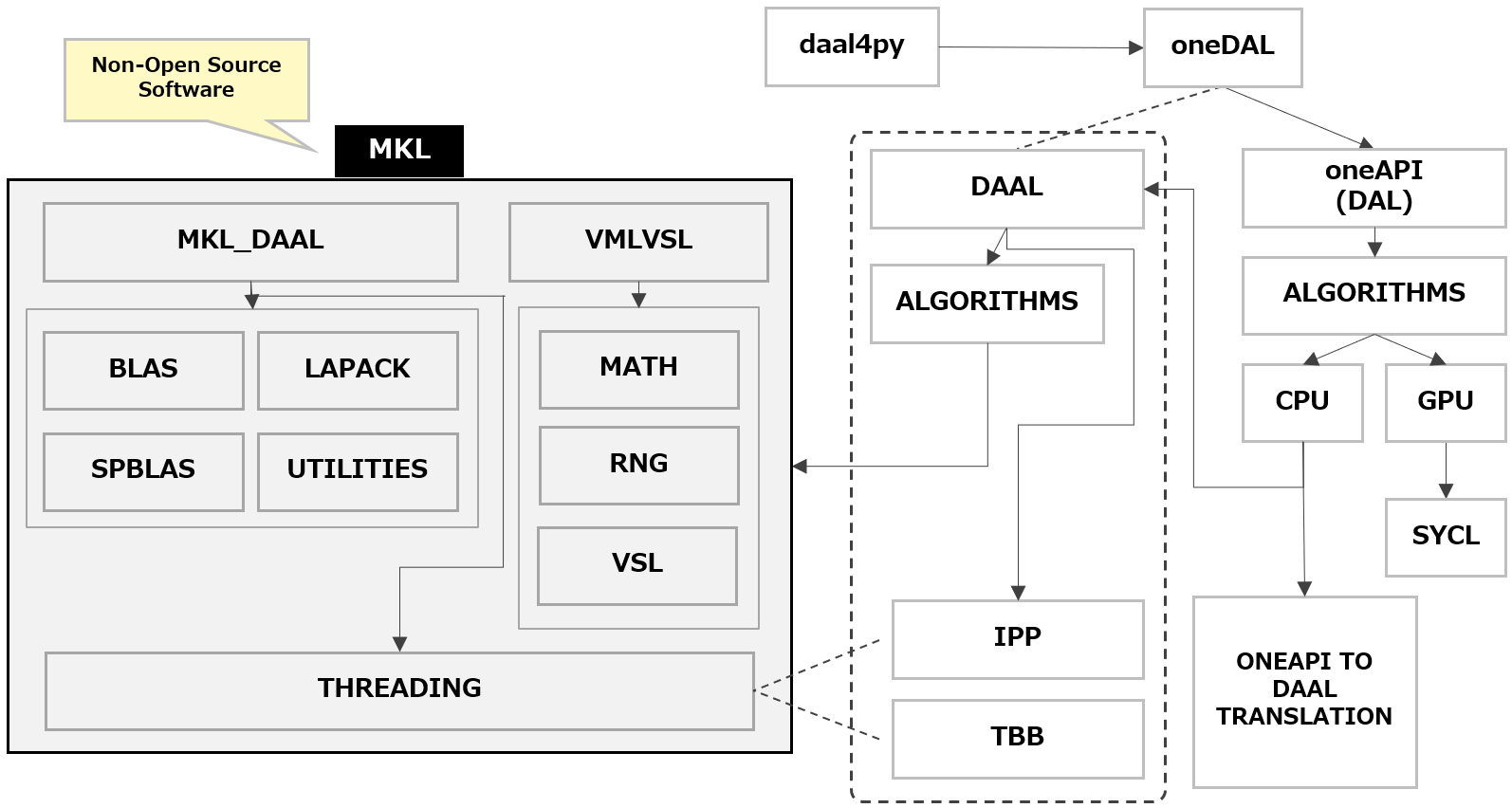} \caption{Dependencies of oneDAL on MKL} \label{fig:mkl} \end{figure}

Efforts to replace MKL with alternatives like OpenBLAS have gained traction to broaden oneDAL’s compatibility with ARM architectures. OpenBLAS provides cross-platform functionality with optimizations for processors such as ARM, RISC-V, sandybridge \cite{sandybridge} and Loongson \cite{loongson}. However, OpenBLAS lacks advanced modules like Sparse BLAS (SPBLAS) and the Vector Statistical Library (VSL) offered by MKL, which are essential for high-performance ML workloads (see Figure~\ref{fig:mkl}) \cite{gotoblaswiki, openblaswiki}. These developments hold the potential to further enhance oneDAL’s high-performance capabilities across diverse hardware ecosystems.


MKL’s advanced capabilities, including SPBLAS for sparse matrix operations and VSL for statistical computations, underpin its exceptional performance in x86-based ML applications. VSL supports a wide range of pseudo-random number generators (RNG) and vectorized statistical routines, critical for ML tasks. Additionally, MKL ensures efficient multi-threaded performance with libraries like IPP (Integrated Performance Primitives) and TBB (Threading Building Blocks) \cite{tbbipp}, which facilitate large-scale parallel computations. In contrast, while OpenBLAS offers essential BLAS and LAPACK functionalities, it lacks the advanced capabilities of SPBLAS and VSL. As a result, OpenBLAS often lags behind MKL in tasks such as sparse matrix operations and compact GEMM (General Matrix Multiply) operations. Despite its open-source flexibility, OpenBLAS performance for these advanced tasks is not yet on par with MKL. However, ongoing developments, such as the addition of support for Tall-and-Skinny Matrix Multiplication (TSMM) \cite{tsmm}, are helping to bridge this gap and improve its applicability in deep learning. These developments hold the promise of further enhancing oneDAL’s performance across diverse hardware ecosystems.

\textbf{Sparse Matrix Processing for ARM Architectures}
Sparse matrix operations are fundamental in machine learning (ML) workloads, especially when processing large datasets with sparse structures containing numerous zero elements. Efficient handling of sparse matrices is critical for algorithms such as Principal Component Analysis (PCA), KMeans clustering, and linear regression, which depend heavily on matrix-vector and matrix-matrix operations. Traditionally, x86 platforms benefit from oneAPI Math Kernel Library, which includes the Sparse BLAS (SPBLAS) module—a robust collection of routines tailored to accelerate these computations. MKL's SPBLAS \cite{mklspblas} module provides a comprehensive set of operations for sparse matrices and vectors, organized into four main groups:

\begin{itemize}
    \item \textbf{State Management Routines:} These routines initialize, configure, and manage data structures, such as \texttt{sparse::matrix\_handle\_t}, which encapsulates sparse matrix representations.
    \item \textbf{Analysis Routines:} Often referred to as the "inspector" or "optimize" stage, these routines inspect matrix properties (e.g., size, sparsity pattern, and parallelism). They create optimized internal data structures or apply transformations to enable efficient execution, without altering user-provided data.
    \item \textbf{Execution Routines:} These perform the actual matrix-vector and matrix-matrix operations, leveraging optimizations and metadata from the analysis stage.
    \item \textbf{Helper Routines:} These utilities handle data manipulation tasks, such as transposing matrices, copying data between handles, or reformatting sparse matrices.
\end{itemize}

This modular and highly optimized structure has enabled x86 systems to efficiently process sparse data, achieving significant performance gains in ML and data analytics workloads. However, OpenBLAS, a widely used alternative to MKL for non-x86 platforms, does not offer equivalent support for sparse computations. This creates a performance gap when running sparse matrix operations on ARM systems. To bridge this gap, our work focuses on porting and optimizing sparse BLAS routines for ARM architectures within the oneDAL framework. These efforts involve developing tailored optimizations to efficiently process sparse matrices in formats such as Compressed Sparse Row (CSR). 



\textbf{Vector Statistical Library (VSL) for ARM Architectures}
Statistical functions are critical to numerous machine learning and data analysis workflows, supporting tasks such as feature scaling, dimensionality reduction, and model evaluation. MKL includes the VSL, which provides highly optimized implementations for x86 platforms. VSL offers advanced functionalities such as random number generation, convolution and correlation, and summary statistics calculations, all structured around task objects data descriptors that encapsulate operation parameters, enabling efficient computation and reusability\cite{mklvsl}. These functions are vital for handling large-scale datasets and computationally intensive ML workflows, including variance calculations, cross-products, and covariance matrices. However, ARM platforms have historically lacked equivalent support, creating performance bottlenecks in statistical computations. To address this gap, this work adapts two key VSL routines, variance calculation \texttt{x2c\_mom} and cross-product computation \texttt{xcp} for ARM architectures. These operations are essential in preprocessing and ML algorithms such as PCA and linear regression.

\textbf{Random Number Generation (RNG) for ARM Architectures}

Random Number Generation plays a crucial role in many machine learning algorithms. In x86-based workflows, oneDAL leverages MKL Vector Statistical Library RNG, which supports a wide range of random number engines and distributions. This includes pseudorandom, quasi-random, and non-deterministic random number generation, providing efficient and scalable performance on x86 platforms. However, on ARM platforms, RNG functionality has been limited to the standard C++ RNG backend, which offers only basic engines, such as MT19937 (Mersenne Twister). This limitation has hindered the performance and versatility of ARM-based ML workflows, particularly for large-scale and parallelized applications.

To overcome this limitation, OpenRNG was integrated as the RNG backend for ARM optimized oneDAL. OpenRNG, an open-source library, was originally released with Arm Performance Libraries 24.04 \cite{armpl} and is designed as a replacement for MKL VSL RNG. It enhances the RNG capabilities on ARM by supporting advanced engines like MT19937 and MCG59 (Multiplicative Congruential Generator). OpenRNG is engineered to maximize performance in multi-threaded and distributed environments, providing three parallel generation methods:
\begin{enumerate}
    \item \textbf{Family Method}: Generates independent streams of random numbers across threads.
    \item \textbf{SkipAhead Method}: Skips forward in the random sequence to generate disjoint subsets.
    \item \textbf{LeapFrog Method}: Allocates alternating elements of the sequence to different threads for parallelism.
\end{enumerate}

OpenRNG’s integration has been shown to provide substantial performance improvements for other experiments such as 44x speedup for PyTorch’s dropout layer and 2.7x speedup over the C++ standard library RNG \cite{openrng}. This makes OpenRNG a key enabler for accelerating machine learning workflows on ARM platforms, particularly in applications requiring high-quality randomness, such as stochastic modeling and AI-driven simulations.

\textbf{SVM Algorithms for ARM SVE Architectures}
Support Vector Machines (SVMs) are widely employed in machine learning (ML) tasks, including classification and regression. In this work, the SVM implementation within oneDAL has been optimized for ARM SVE, enhancing computational throughput and enabling efficient parallel execution. Initially, the SVM performance on ARM was comparable to or worse than its x86 using the MKL backend counterpart. This performance gap prompted a deeper analysis of the implementation to optimize and ensure that the ARM SVE optimized oneDAL could match or exceed the performance of default oneDAL on x86 with MKL. Upon further investigation, we identified key computational bottlenecks that could be alleviated through fine-tuning and optimization using SVE’s unique capabilities. Key computational processes within SVM, such as working set selection (WSS) and Lagrange multiplier updates, were re-engineered to leverage SVE's features. The vector-length agnostic (VLA) processing of SVE allows dynamic adjustment to varying hardware configurations, while predicate registers ensure efficient handling of conditional computations by enabling selective data processing within vectorized instructions.

\section{Contributions}

This work makes several important contributions to the adaptation and optimization of the oneDAL library for ARM processors:

\begin{enumerate}
    \item The oneDAL was successfully ported to ARM SVE and NEON architecture by utilizing OpenBLAS alternative of MKL, enabling high-performance machine learning and data analytics on ARM-based systems while maintaining compatibility with x86 architectures.

    \item Efficient sparse matrix operations were implemented within oneDAL, specifically designed for compressed sparse row (CSR) format data. The newly developed routines, including \texttt{csrmm}, \texttt{csrmultd}, and \texttt{csrmv}, enable oneDAL to perform sparse matrix computations effectively on ARM processors.

    \item A novel implementation of the Vector Statistical Library (VSL) was developed, enabling statistical routines, such as variance calculations and cross-product matrix computations, to be executed efficiently on ARM hardware.
    
    \item OpenRNG was integrated as the backend random number generator for oneDAL, replacing the default stdc++ implementation on non-x86 platforms. This integration extended the RNG functionalities available in oneDAL, opening up more optimization opportunities.

    \item The Support Vector Machine (SVM) algorithm was optimized, with a specific focus on the Working Set Selection (WSS) function, to take advantage of ARM’s SVE features. By utilizing SVE's vectorization capabilities, significant performance improvements were achieved on ARM-SVE based systems.

\end{enumerate}
The optimizations and validations implemented are publicly available as part of the \textit{oneDAL} open-source project on GitHub \cite{onedalPR}, allowing the community to use these enhancements for ARM support.

\section{Methodology and Implementation}

This section presents a structured approach to enabling \textit{oneDAL} for ARM architectures, with a focus on ARM’s Scalable Vector Extension (SVE) optimizations. Key areas covered include ARM enablement, Sparse BLAS optimizations, Vector Statistical Library (VSL) implementations, Support Vector Machine (SVM) optimizations with Scalable Vector Extension, and Random Number Generator (OpenRNG) integration. Each subsection is demonstrating how these optimizations collectively enhance \textit{oneDAL’s} on ARM SVE Architecture.

\subsection{Optimizing oneDAL for ARM SVE Compatibility}

The primary objective of optimization is to adapt \textit{oneDAL} to work efficiently on ARM platforms without dependency on Math Kernel Library (MKL). To achieve this, we utilized OpenBLAS BLAS library optimized for ARM and compatible with Scalable Vector Extension (SVE).

\begin{figure}[h]
    \centering
    \includegraphics[width=1\linewidth]{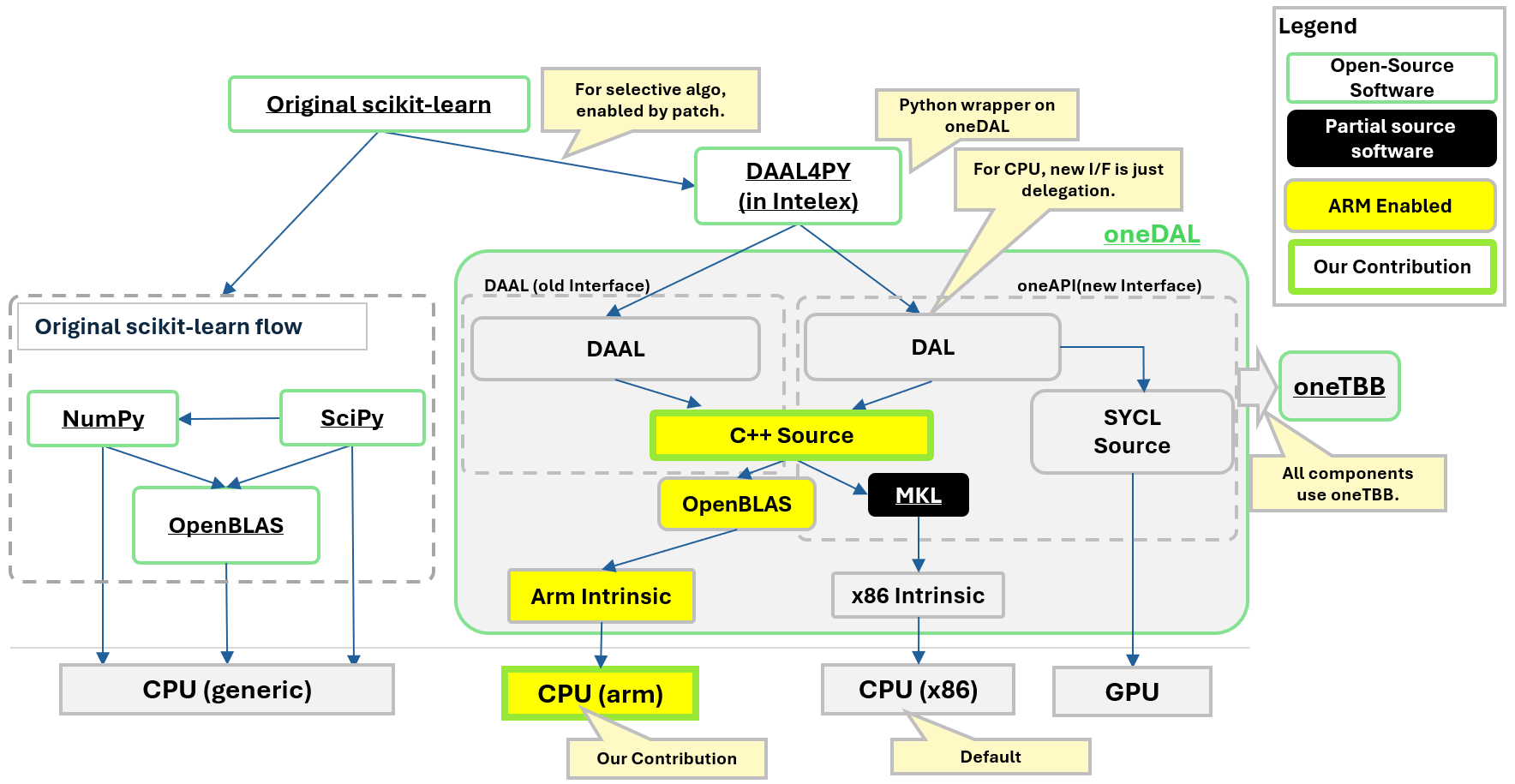}
    \caption{Proposed Architecture: ARM SVE optimized oneDAL}
    \label{fig:method}
\end{figure}

Figure \ref{fig:method} illustrates our contribution to enable oneDAL for ARM architecture.

The stages for ARM enablement and adoption of OpenBLAS as a computational engine include:
\begin{itemize}
    \item \textbf{Introduction of New Template}: This research incorporated new template parameters and macros specific to ARM architectures as addition to existing architectures. Which allow more tailored and efficient execution of algorithms on ARM hardware without disrupting existing oneDAL functionality and architectures support.
    \item \textbf{Refactoring Kernel Definitions and ARM-Specific Templates}: Architecture-specific kernel definitions were refactored, and ARM-specific template parameters were introduced to improve code organization and maintainability. These changes allow \textit{oneDAL} to run efficiently on both ARM and x86 architectures without compromising code integrity. The refactored structure makes it easier to distinguish between ARM and x86-specific implementations, while ensuring that algorithms are tailored for ARM hardware.

    \item \textbf{Dynamic CPU Dispatch Mechanism}: To handle diverse ARM configurations, we implemented a CPU dispatch mechanism that dynamically selects the most appropriate vectorized instructions (NEON or SVE) based on the ARM CPU’s capabilities. This mechanism ensures that \textit{oneDAL} utilizes the most efficient vectorization approach available on the target hardware.

    \item \textbf{Conditional Compilation}: ARM-specific optimizations were integrated using conditional compilation, which isolates ARM code paths from x86 paths. Compiler macros enable selective inclusion of ARM-specific code only when compiling for ARM.
    \begin{verbatim}
        #ifdef __ARM_SVE
        // SVE-optimized code
        #else
        // Other_Architecture code
        #endif
    \end{verbatim}

    \item \textbf{Tailored Build System}: The build system was adapted to detect the target architecture and apply appropriate compiler flags. ARM-specific flags activate SVE.
\end{itemize}

This comprehensive approach enabled the successful adaptation of \textit{oneDAL} for ARM, ensuring cross-platform compatibility. With the core ARM enablement in place, we shifted our focus towards optimizing key computational routines and expanding the support for additional algorithms on the ARM platform.

\subsection{Sparse BLAS Implementations}

Sparse matrix operations are essential in high-performance computing, particularly in machine learning (ML) and data analytics workloads, where large and sparse datasets are common. In \textit{oneDAL}, three key routines are required to handle data in the Compressed Sparse Row (CSR) format: \texttt{csrmm}, \texttt{csrmultd}, and \texttt{csrmv}. Unlike BLAS and LAPACK standards, no universally accepted standard exists for sparse matrix operations \cite{openblasfaq}. Traditionally, \textit{oneDAL} relies on MKL for these operations. However, it cannot be ported directly to ARM. To address this, we developed reference C++ implementations for these sparse routines based on MKL’s functionality specifications, enabling compatibility with ARM and other platforms. This implementation is crucial for enabling algorithms such as PCA, covariance, correlation, and KMeans on ARM-based systems.

The sparse routines required by \textit{oneDAL} are defined as follows:
\begin{enumerate}
    \item \textbf{csrmm}: $C\leftarrow \alpha \hbox{op}(A)B + \beta C$, where A is a sparse matrix in CSR format and B, C are dense matrices.
    \item \textbf{csrmultd}: $C:=\hbox{op}(A)B$, where A and B are sparse matrices in the CSR format and C is a dense matrix.
    \item \textbf{csrmv}: $y\leftarrow \alpha \hbox{op}(A)x + \beta y$, where A is a sparse matrix in the CSR format and x, y are densely formatted vectors.
\end{enumerate}
Here $op$ is either the identity or transpose operation.

The function \textit{csrmm} already had a reference implementation. So, following will focus on the implementation details of the other two.

\subsubsection{csrmultd}
This routine requires $A$ and $B$ in 3-array CSR form, with 1-based indexing on the index arrays. Matrix C is dense and is stored in column-major format. The routine supports both $AB$ and $A^tB$, therefore separate compute kernels for these operations were implemented.

For $AB$, the $(i,j)$th entry of C is obtained by iterating the below method over $k$. 
$$ C_{ij}\leftarrow A_{ik}B_{kj} $$ 

The entire matrix is obtained by iterating over $i$ and $j$. The ideal order would be a row-wise iteration over $A$ and $B$ (because they are stored in CSR format); and a column-wise iteration over $C$ (because it is stored in column major order). However, the column traversal of $C$ and the row traversal of $A$ cannot be achieved at the same time, since they share the same column index ($i$). 

Therefore, this work has to choose between the following options: 
\begin{enumerate}
    \item[a).] Row traversal on $A$ and column traversal on $C$; 
    \item[b).] Column traversal on $A$ and row traversal on $C$. 
\end{enumerate}

CSR format makes the second option complicated, so this work decided to use the former. The nested loops are in the order $j$--$k$--$i$ (innermost to outermost). For $A^tB$, the $(i,j)$th entry of C is obtained by iterating $$ C_{ij}\leftarrow A_{ki}B_{kj} $$ over $i, j, k$. Here, the ideal order of iterations, which is column-wise traversal over $C$ and row-wise traversal over $A$ and $B$, can be achieved by using $i$--$j$--$k$ (inner to outer) as the order of iteration. This work implementation does the same.

\subsubsection{csrmv}
This routine requires $A$ in 4-array CSR form. The index arrays can be either zero-based or one-based. Similarly to the previous case, separate kernels were implemented for $Ax+y$ and $A^tx+y$. The choice of the order of iteration is obvious in this case because there are only two choices, of which this work chooses the one that involves a row-order traversal of $A$. 

Initial performance evaluations indicate that while these implementations do not yet match the speed of MKL, they provide essential sparse matrix functionality for ARM architectures. These implementations establish a foundation for future optimizations, targeting ARM-specific architectural features to improve efficiency and close the performance gap with MKL.

\subsection{Vector Statistical Library (VSL) Implementations}

The \textit{oneDAL} library's Vector Statistical Library (VSL) traditionally relies on MKL for performing statistical computations, which has restricted its functionality on ARM-based platforms and led to compilation challenges for certain algorithms. To overcome these limitations, ARM-compatible implementations of essential VSL routines were developed. This section highlights the ARM-optimized implementations of two key routines: \texttt{x2c\_mom}, used for variance calculations, and \texttt{xcp}, for computing cross-products. These enhancements enable \textit{oneDAL} to execute statistical operations independently of MKL, fully utilizing ARM's architecture.

\subsubsection{Variance Calculation (\texttt{x2c\_mom})}

The \texttt{x2c\_mom} kernel computes the sample variance across each coordinate for a dataset matrix \( X \in \mathbb{R}^{p \times n} \), where each column represents a \( p \)-dimensional sample. The variance \( v_i \) for the \( i \)-th coordinate is defined as:
\begin{equation}
    v_i = \frac{1}{n - 1} \sum_{j=1}^n (X_{ij} - \mu_i)^2
\end{equation}
where \( \mu_i \), the sample mean of the \( i \)-th coordinate, is calculated by:
\begin{equation}
    \mu_i = \frac{1}{n} \sum_{j=1}^n X_{ij}.
\end{equation}

To enhance performance, this expression was reformulated using the first and second raw moments, \( S^{(1)}_i \) and \( S^{(2)}_i \), as follows:
\begin{equation}
    v_i = \frac{S_i^{(2)}}{n - 1} - \frac{(S_i^{(1)})^2}{n(n - 1)}
\end{equation}
where:
\[
S^{(1)}_i = \sum_{j=1}^n X_{ij} \quad \text{and} \quad S^{(2)}_i = \sum_{j=1}^n X_{ij}^2.
\]
This formulation minimizes recalculations of means, thus facilitating vectorization and parallel execution using ARM SVE.

\subsubsection{Matrix of Cross Products (\texttt{xcp})}

The \texttt{xcp} kernel calculates the cross-product matrix \( C \in \mathbb{R}^{p \times p} \) for a dataset \( X \in \mathbb{R}^{p \times n} \). Each element \( C_{ij} \) of this matrix is defined as:

\begin{equation}
    C_{ij} = \sum_{k=1}^n (X_{ik} - \mu_i)(X_{jk} - \mu_j)
\end{equation}
where \( \mu_i \) and \( \mu_j \) are the means along coordinates \( i \) and \( j \). 

The routine supports batch-wise computation, where the cross-product matrix, the sum, and the number of observations from the previous batch are passed to the function. The function then updates the cross-product matrix to reflect the combined data. For data processed in two batches, with \( 1, \ldots, n' \) observations in the first batch and \( n'+1, \ldots, n'+n \) in the second, the cross-product matrix for each element \( C_{ij} \) is calculated as:

\begin{equation}
\begin{aligned}
    C_{ij} &= \sum_{k=1}^{n'} \big((X_{ik} - \mu'_i) + (\mu'_i - \mu_i)\big) \\
    &\quad \times \big((X_{jk} - \mu'_j) + (\mu'_j - \mu_j)\big) \\
    &\quad + \sum_{k=n'+1}^{n'+n} (X_{ik} - \mu_i)(X_{jk} - \mu_j).
\end{aligned}
\end{equation}

This expression is optimized by using:
\begin{equation}
    C \leftarrow C' + \frac{S' (S')^T}{n'} - \frac{S S^T}{n} + X X^T,
\end{equation}
where \( S' \) is the raw sum of the first batch, \( S \) the cumulative raw sum, and \( C' \) the previously computed cross-product matrix. Leveraging BLAS routines, this formula allows for memory-efficient computation, further improved by ARM SVE-based parallel processing.

This implementation delivers two new VSL implementation to open source community over traditional scalar approaches. These optimizations establish \textit{oneDAL} as a viable library for data-intensive applications on ARM platforms.
\subsection{Random Number Generation (RNG) Optimization}

On ARM-based architectures, \textit{oneDAL} faced limitations in random number generation (RNG) due to its reliance on the \texttt{stdc++} implementation, which lacks advanced RNG capabilities critical for data science and machine learning workflows. To bridge this gap and achieve feature parity across platforms, we integrated \texttt{OpenRNG} as the RNG backend for \textit{oneDAL} on ARM. \texttt{OpenRNG} is an open-source library offering a robust set of RNG functionalities, designed to closely replicate the interface and performance of MKL RNG, while incorporating optimized kernels tailored for high-performance computation.

\subsubsection{Integration of OpenRNG into oneDAL}

The integration of \texttt{OpenRNG} required several modifications to the \textit{oneDAL} codebase to replace the default \texttt{stdc++} RNG backend. This involved creating a new header file, \texttt{service\_rng\_openrng.h}, which facilitates the interaction between \textit{oneDAL} and \texttt{OpenRNG}. Since the \texttt{OpenRNG} interface is highly compatible with the MKL RNG interface, minimal changes were required in the existing RNG service files, specifically in \texttt{service\_stat\_mkl.h}. 

Key modifications included:
\begin{itemize}
    \item Developing a new \texttt{service\_rng\_openrng.h} header file to define the interface for \texttt{OpenRNG}.
    \item Adjusting \texttt{service\_stat\_mkl.h} to accommodate the interface similarities between MKL RNG and \texttt{OpenRNG}, thereby enabling seamless integration.
    \item Updating build configuration files to enable static linking with the \texttt{OpenRNG} library, ensuring that the advanced RNG functionalities are available in the final build across all supported hardware platforms.
\end{itemize}

\subsubsection{Enhanced RNG Functionality and Performance Implications}

The integration of \texttt{OpenRNG} into \textit{oneDAL} introduces an expanded set of RNG capabilities on ARM platforms. Although RNG represents a relatively minor part of the overall computational workload, the integration of \texttt{OpenRNG} allows \textit{oneDAL} to maintain functionality parity across platforms.

\begin{figure}[h]
    \centering
    \includegraphics[width=0.9\linewidth]{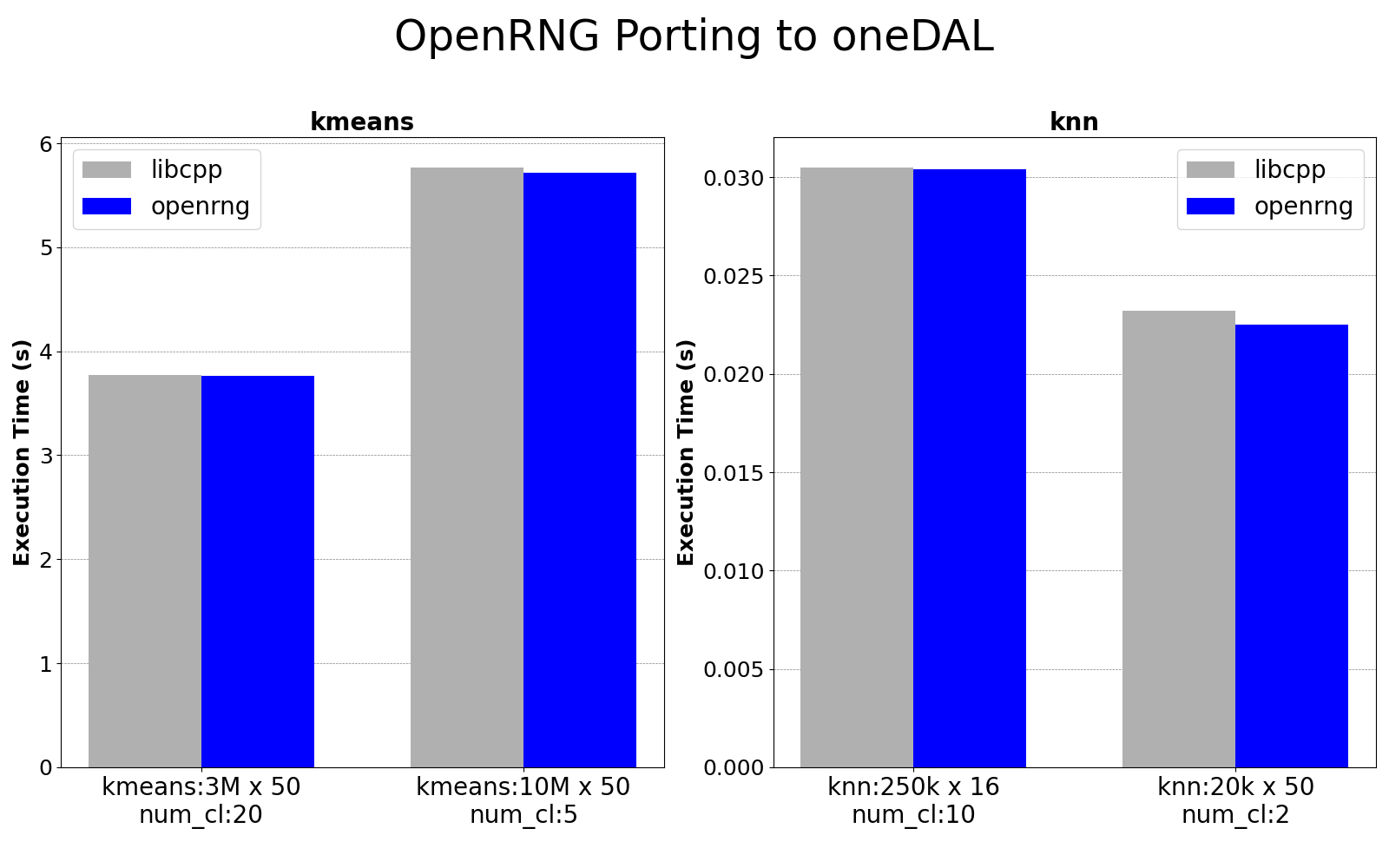}
    \caption{Performance of KNN and KMeans with \texttt{libcpp} vs. \texttt{OpenRNG}}
    \label{fig:rng_benchmark}
\end{figure}

\begin{itemize}
    \item \textbf{Support for Multiple RNG Engines:} \texttt{OpenRNG} conforms to the MKL VSL RNG specification, supporting engines such as \texttt{MT19937} and \texttt{mcg59}. Currently, \texttt{mt2203} is not included in \texttt{OpenRNG}, but adding it could further improve performance for algorithms like Random Forests on ARM. In comparison, \texttt{stdc++} only supports the \texttt{MT19937} engine.
    \item \textbf{Optional Backend Configuration:} Users can now select their preferred RNG backend at compile time, allowing flexibility between \texttt{OpenRNG} on ARM and MKL on x86, thereby promoting cross-platform consistency with tailored optimizations for each architecture.
    \item \textbf{Performance Parity and Cross-Platform Consistency:} Benchmark results Figure \ref{fig:rng_benchmark} indicate that while RNG contributes minimally to the total workload time, \texttt{OpenRNG} on ARM provides comparable functionality to MKL on x86, achieving uniform RNG performance across both ARM and x86 systems.
\end{itemize}

Figure \ref{fig:rng_benchmark} highlights the performance of RNG-dependent algorithms such as KNN and KMeans using the \texttt{libcpp} and \texttt{OpenRNG} backends, demonstrating that the adoption of \texttt{OpenRNG} maintains competitive performance without incurring additional overhead. By integrating \texttt{OpenRNG}, \textit{oneDAL} gains a robust and consistent RNG backend for ARM architectures, enhancing its support for high-performance, ARM-optimized data analytics workflows while ensuring consistent performance and reproducibility across diverse hardware platforms.

\subsection{SVM Optimization for ARM optimized oneDAL Using SVE}

Support Vector Machines (SVM) algorithms can be computationally intensive, especially with large datasets, due to the increased number of constraints that need to be optimized to maximize the margin of separation. The oneDAL library provides an accelerated version of SVM, utilizing a Working Set Selection (WSS) mechanism based on second-order information to handle this complexity efficiently. Specifically, oneDAL uses the WSS3 variant, where pairs of indices are selected in each iteration for optimizing the Lagrange multipliers.

In this work, we focus on optimizing the `WSSj` function in oneDAL's SVM implementation for ARM architectures with Scalable Vector Extension (SVE). The `WSSj` function selects the index \( j \) to update the Lagrange multipliers. Although compilers typically vectorize loops, the data dependencies in `WSSj` hinder automatic vectorization. ARM’s SVE, with its predicate capabilities, offers an effective solution to this issue, enabling conditional operations within vectorized loops.

\subsubsection{Original Loop in \texttt{WSSj} Function}

The original scalar loop in the `WSSj` function iterates over a range defined by \( j_{\text{start}} \) to \( j_{\text{end}} \). Multiple \texttt{if} conditions filter values based on specific criteria, which complicates automatic vectorization. Below is the original loop implementation in C++.

\begin{lstlisting}[language=C++, caption=Original Loop in \texttt{WSSj} Function]
for (size_t j = jStart; j < jEnd; j++) {
    const algorithmFPType gradj = grad[j];
    if (!(I[j] & sign)) {
        continue;
    }
    if ((I[j] & low) != low) {
        continue;
    }
    if (gradj > GMax2) {
        GMax2 = gradj;
    }
    if (gradj < GMin) {
        continue;
    }
    const algorithmFPType b = GMin - gradj;
    algorithmFPType a = Kii + kernelDiag[j] - two * KiBlock[j - jStart];
    if (a <= zero) {
        a = tau;
    }
    const algorithmFPType dt = b / a;
    const algorithmFPType objFunc = b * dt;
    if (objFunc > GMax) {
        GMax = objFunc;
        Bj = j;
        delta = -dt;
    }
}
\end{lstlisting}

\subsubsection{SVE-Optimized Loop in \texttt{WSSj} Function}

To overcome the vectorization limitation, we implemented an SVE optimized version of the loop using predicates. ARM SVE predicates allow for conditionally executing operations within vectorized loops, thus handling the multiple \texttt{if} conditions in a single pass. The optimized loop leverages SVE's vector length adaptability, ensuring compatibility across ARM systems with different vector widths.

\begin{lstlisting}[language=C++, caption=SVE-Optimized Loop in \texttt{WSSj} Function]
for (size_t j_cur = jStart; j_cur < jEnd; j_cur += w) {
    svint32_t Bj_vec_cur = svindex_s32(j_cur, 1);  // Vectorized index for Bj
    svbool_t pg2 = svwhilelt_b32(j_cur, jEnd);     // Predicate for vector length adaptation

    svint32_t vec_I = svld1sb_s32(pg2, reinterpret_cast<const int8_t *>(&I[j_cur])); // Load condition vector
    
    // Combine the `if` conditions
    svint32_t result_of_and32 = svand_s32_m(pg2, vec_I, vecSignLow);
    pg2 = svcmpeq_s32(pg2, result_of_and32, vecSignLow); // Predicate-based filtering
    
    // Remaining vectorized logic...
}
\end{lstlisting}

In this SVE-optimized loop:
\begin{itemize}
    \item \textbf{Predicate Initialization}: The variable \texttt{pg2} adapts to the available vector length, allowing the code to run efficiently on ARM hardware with varying vector lengths.
    \item \textbf{Combined Conditions}: The \texttt{if} conditions \texttt{!(I[j] \& sign)} and \texttt{(I[j] \& low) != low} are merged using predicates, enabling selective processing within the vectorized loop.
    \item \textbf{Vectorized Execution}: The loop executes vectorized operations across multiple indices simultaneously, dynamically adjusting to the hardware’s vector length.
\end{itemize}

\subsubsection{Performance Gains}

\begin{figure}[h]
    \centering
    \includegraphics[width=0.9\linewidth]{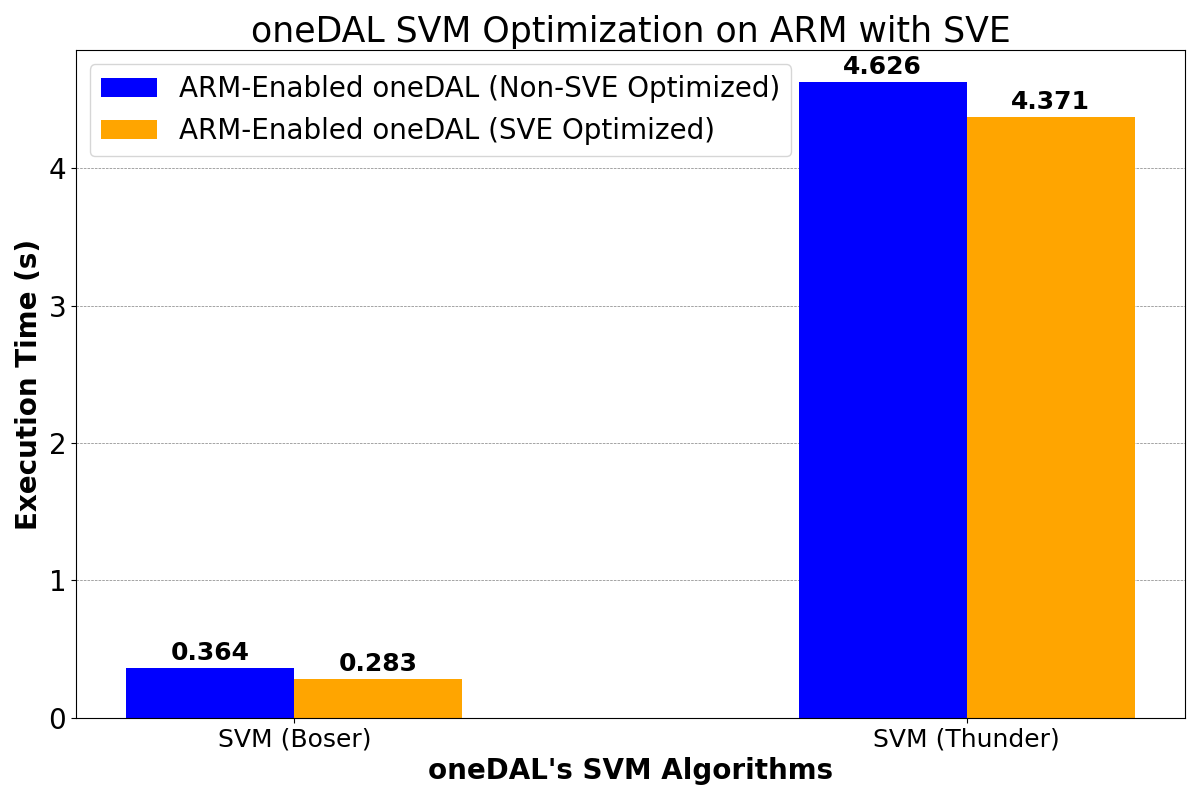}
    \caption{Performance of SVM Non-SVE vs. SVE Optimized}
    \label{fig:svmsve}
\end{figure}

After implementing the SVE vectorized version of the \texttt{WSSj} function, we achieved substantial performance improvements shown in Figure \ref{fig:svmsve}.
\begin{itemize}
    \item The \textbf{Boser method} of the SVM algorithm demonstrated a \textbf{22\% performance gain}.
    \item The \textbf{Thunder method} exhibited a \textbf{5\% performance gain}.
\end{itemize}
These performance gains were achieved on AWS Graviton3 instances, which feature an SVE vector length of 256 bits, under a single-core configuration. The optimized vectorized loop maintained bitwise accuracy compared to its scalar base implementation, ensuring the fidelity of results. The successful vectorization of \texttt{WSSj} and the associated performance improvements highlight the potential of ARM SVE in enhancing SVM computations within oneDAL on ARM platforms.

\subsection{Experimental Setup}

\subsubsection{Hardware Specifications}

Table \ref{tab:hardware_comparison} presents a comparison of the ARM-based (Graviton3) and x86-based (Ice-Lake) AWS instances used in this experiment to benchmark the performance of the ARM SVE optimized \textit{oneDAL} in result section.

\begin{table}[h]
    \centering
    \caption{Configurations of ARM and x86 AWS Instances}
    \label{tab:hardware_comparison}
    \begin{tabular}{|c|c|c|}
        \hline
        & \textbf{ARM Machine} & \textbf{x86 Machine} \\
        \hline
        \textbf{AWS Instance} & c7g.8xlarge & c6i.8xlarge \\
        \hline
        \textbf{vCPUs} & 32 & 32 \\
        \hline
        \textbf{Processor} & AWS Graviton3 & Intel Xeon 8375C \\
        \hline
        \textbf{Clock Speed} & 2.5 GHz & 3.5 GHz \\
        \hline
        \textbf{Memory} & 32 GB & 64 GB \\
        \hline
        \textbf{Network Bandwidth} & 15 Gbps & 12.5 Gbps \\
        \hline
        \textbf{EBS Bandwidth} & 10 Gbps & 10 Gbps \\
        \hline
        \textbf{Price} & \$0.7853/hr & \$1.36/hr \\
        \hline
    \end{tabular}
\end{table}

\subsubsection{Development Environment}

The development environment for the ARM SVE optimized \textit{oneDAL} comprised:
\begin{itemize}
    \item \textbf{Operating System}: Ubuntu, running on the ARMv8 architecture.
    \item \textbf{Compiler and Libraries}: GCC 13.0, LLVM 17.0, and OpenBLAS 0.3.26.
    \item \textbf{Profiling Tools}: \texttt{perf} on Linux was used for profiling and performance analysis.
\end{itemize}




\section{Results}

The ARM-SVE optimized \textit{oneDAL} was assessed through an extensive series of benchmarks and practical applications to measure its performance on ARM-based systems. This section presents comparative benchmark results, showcasing the performance of ARM SVE optimized \textit{oneDAL} against the original \textit{scikit-learn} on ARM, as well as the x86 \textit{oneDAL} with MKL backend. We utilize intel's scikit-learn benchmarks\cite{intelbenchmark} to cover a range of machine learning models and datasets, including synthetic, real-world, and industry-relevant use cases. Additionally, results from the TPC-AI Benchmark \cite{tcpai}, Data centric benchmarking i.e. DataPerf Selection Speech Challenge \cite{dataperf} hosted by Dataperf MLcommons \cite{mswc}, and a credit card fraud detection scenarios are included to highlight the practical value of our optimizations.

\subsection{Performance of ARM SVE Optimized oneDAL vs. Original scikit-learn on ARM Platform}

\begin{figure}[h]
    \centering
    \includegraphics[width=1\linewidth]{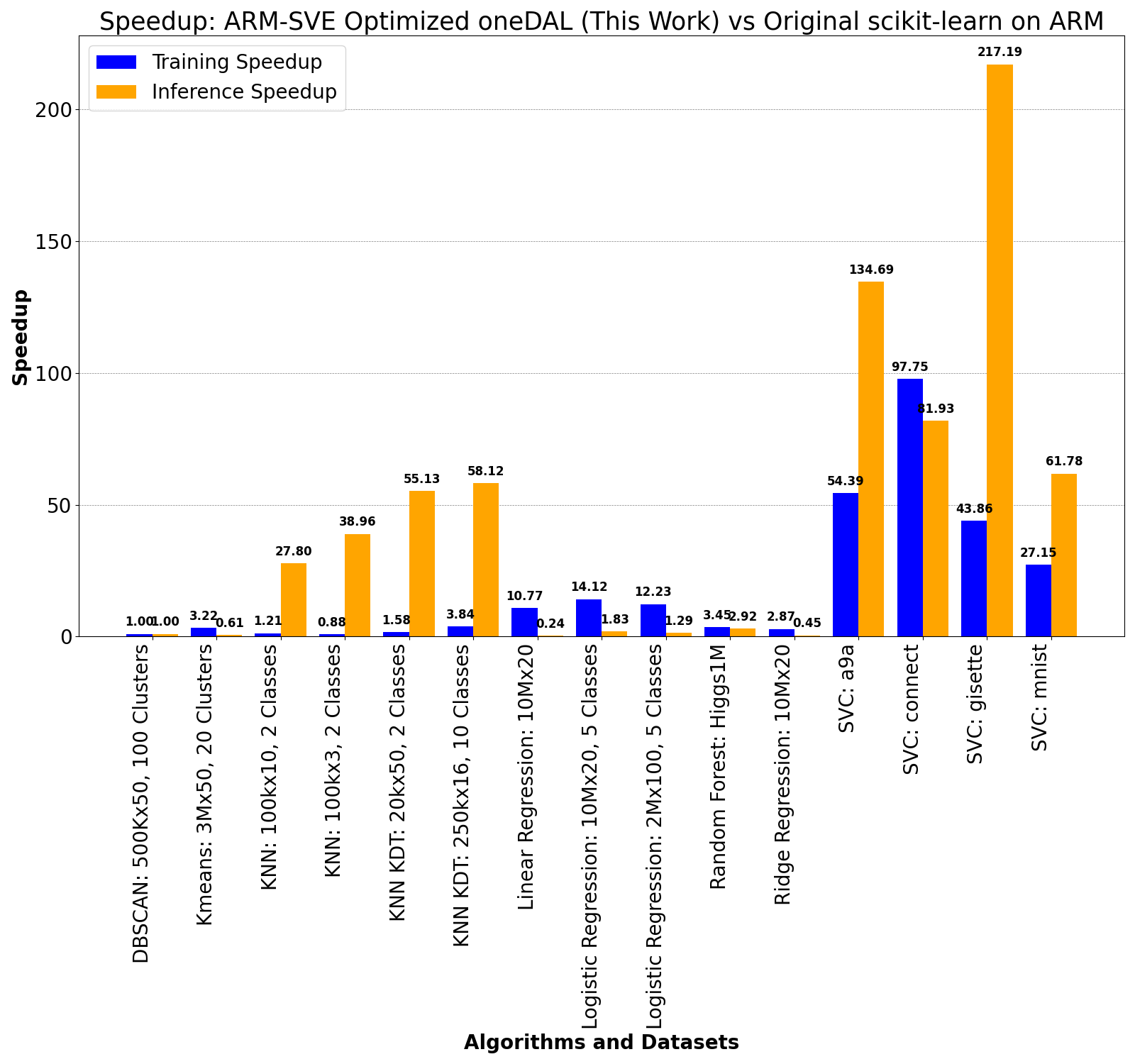}
    \caption{Performance of ARM SVE optimized \textit{oneDAL} vs. original \textit{scikit-learn}}
    \label{fig:armarmtrain}
\end{figure}

The comparison between ARM SVE optimized \textit{oneDAL} and the original \textit{scikit-learn} demonstrates substantial performance gains on ARM architectures, as depicted in Figure \ref{fig:armarmtrain}. The performance improvements were observed across both training and inference tasks, with speedups ranging from 1x to as high as 217x. The most notable speedups were recorded for \textit{SVM} on the \textit{a9 dataset} (134.69x) and \textit{SVM} on the \textit{gisette dataset} (217.19x). These results confirm the significant impact of ARM SVE optimizations on the acceleration of machine learning algorithms.

 However, not all algorithms benefited equally from the optimizations. For example, \textit{DBSCAN} on the \textit{500x3, 100 clusters} dataset showed no significant improvement, with a speedup of just 1.00x during training. This indicates that ARM SVE optimizations are less impactful for density-based clustering tasks, particularly when the dataset has small feature dimensions. Similarly, \textit{Logistic Regression} on the \textit{2Mx100, 5 classes} dataset achieved a modest 1.29x speedup for inference, and linear models such as \textit{Linear Regression} and \textit{Ridge Regression} (both on the \textit{10Mx20} dataset) demonstrated minimal improvements, with speedups of 0.24x (slower) and 0.45x (slower), respectively. These results suggest that further optimization is required in vectorized linear algebra operations to fully exploit the capabilities of ARM SVE. Although our optimizations lead to impressive performance improvements for most algorithms, there are specific cases, particularly with linear models and density-based clustering, where further refinement is needed to achieve maximum efficiency.



\subsection{Performance of ARM SVE Optimized oneDAL vs. x86 oneDAL (MKL)}
The performance comparison of ARM SVE optimized \textit{oneDAL} with x86 oneDAL with MKL backend reveals that the ARM-SVE optimizations deliver competitive speedups in both training and inference across a range of machine learning algorithms and datasets, as shown in Figure \ref{fig:armarmtrain}.

\begin{figure}[h]
    \centering
    \includegraphics[width=1\linewidth]{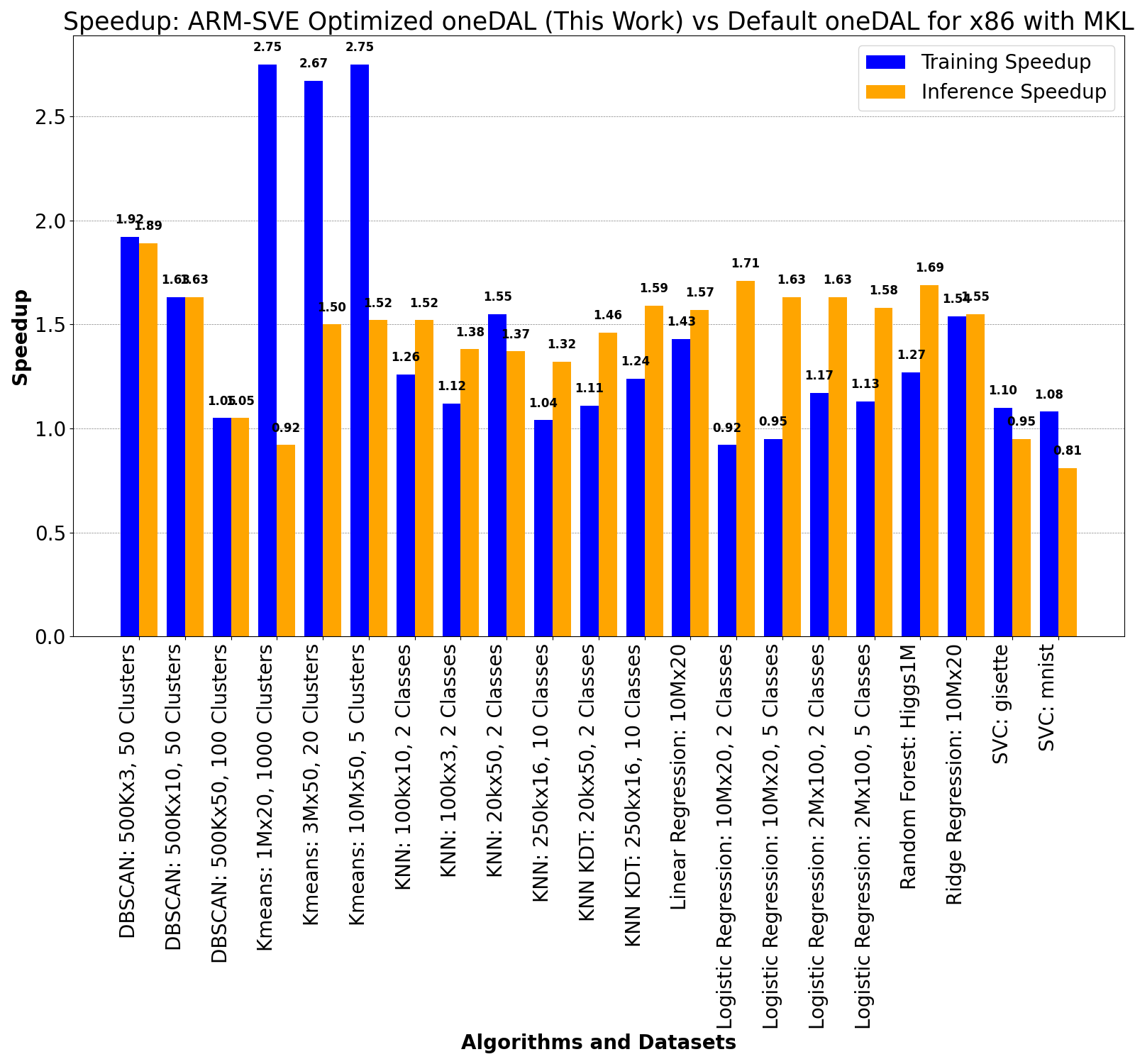}
    \caption{Performance of ARM SVE optimized \textit{oneDAL} vs. x86 oneDAL (MKL)}
    \label{fig:armxtrain}
\end{figure}
In the training phase, our work demonstrates upto 2.75x speedups over the default x86 oneDAL with MKL backend, with the highest improvements seen in KMeans (2.75x) and DBSCAN (1.92x) for clustering workloads. For KNN-based algorithms, our work achieves consistent speedups up to 1.5x, highlighting its effectiveness for distance-based operations. During inference, ARM-SVE optimized oneDAL maintains performance parity or achieves up to 1.83x speedup, particularly excelling in workloads like DBSCAN, Logistic Regression and Linear Regression. Notably, algorithms with high computational complexity, such as SVM (SVC) and Random Forest, achieve comparable performance, further validating the optimization. 





\subsection{Performance Analysis: DataPerf Selection Speech}
\begin{figure}[h]
    \centering
    \includegraphics[width=1\linewidth]{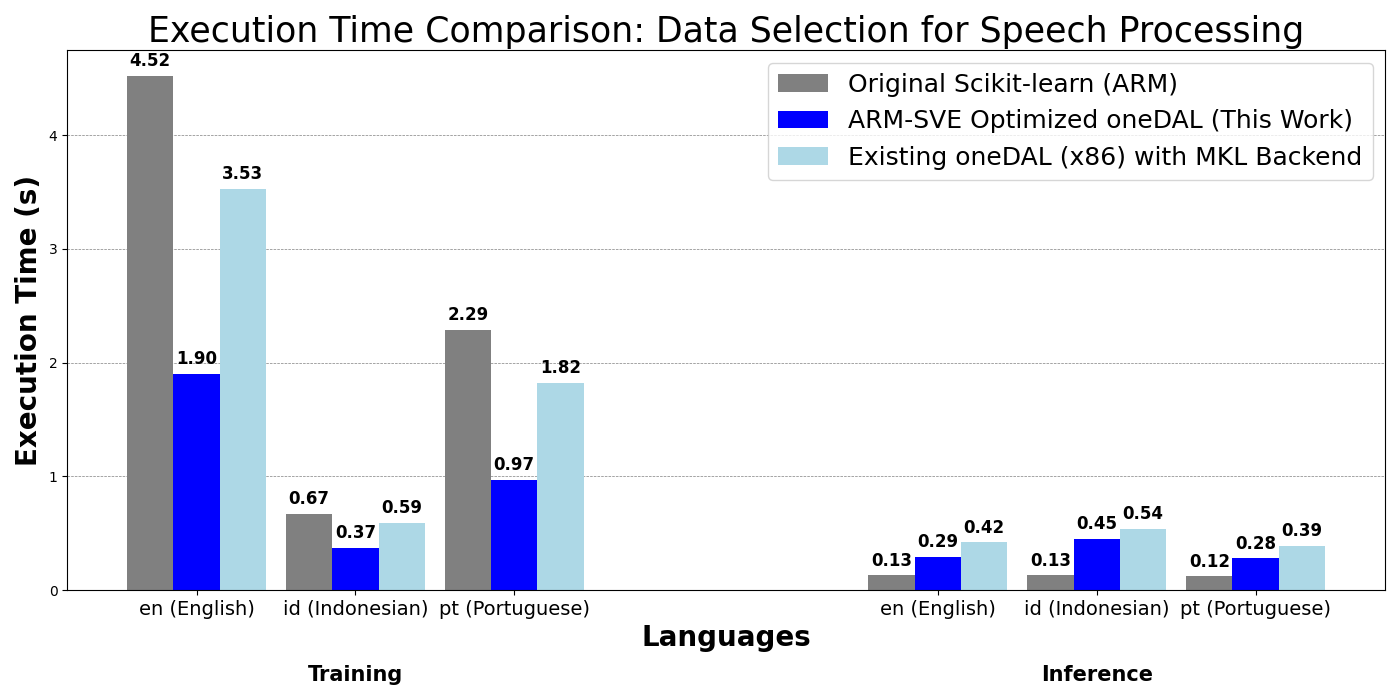}
    \caption{Performance of DataPerf with ARM SVE optimized \textit{oneDAL}.}
    \label{fig:dataperf}
\end{figure}

The DataPerf Selection Speech benchmark evaluates dataset selection algorithms for keyword spotting, focusing on both training and inference execution times across three languages: English (en), Indonesian (id), and Portuguese (pt). The results are shown in Figure \ref{fig:dataperf}, which compares the performance of ARM-SVE optimized oneDAL, x86-based oneDAL with MKL, and original scikit-learn on ARM.

Our work exhibited significant improvements in training times compared to the original scikit-learn, with reductions of 58\% (English), 45\% (Indonesian), and 60\% (Portuguese). This was accompanied by modest gains over the implementation of oneDAL x86 (MKL), ranging from 37\% to 46\%. However, inference performance showed a mixed trend: while our version outperformed x86 oneDAL (MKL) across all languages and but still inference times were higher than original scikit-learn on ARM.

These results underscore the robustness of ARM-SVE optimizations oneDAL in accelerating data selection challenges, particularly in training, while narrowing the performance gap in inference.

\subsection{Performance Analysis: TPC-AI Benchmark}
The TPC-AI benchmark evaluates the end-to-end performance of machine learning workloads relevant to industry AI applications. We focused on the customer segmentation use case, which uses K-means clustering on a 1GB synthetic dataset \cite{tcpai}.

\begin{figure}[h]
    \centering
    \includegraphics[width=1\linewidth]{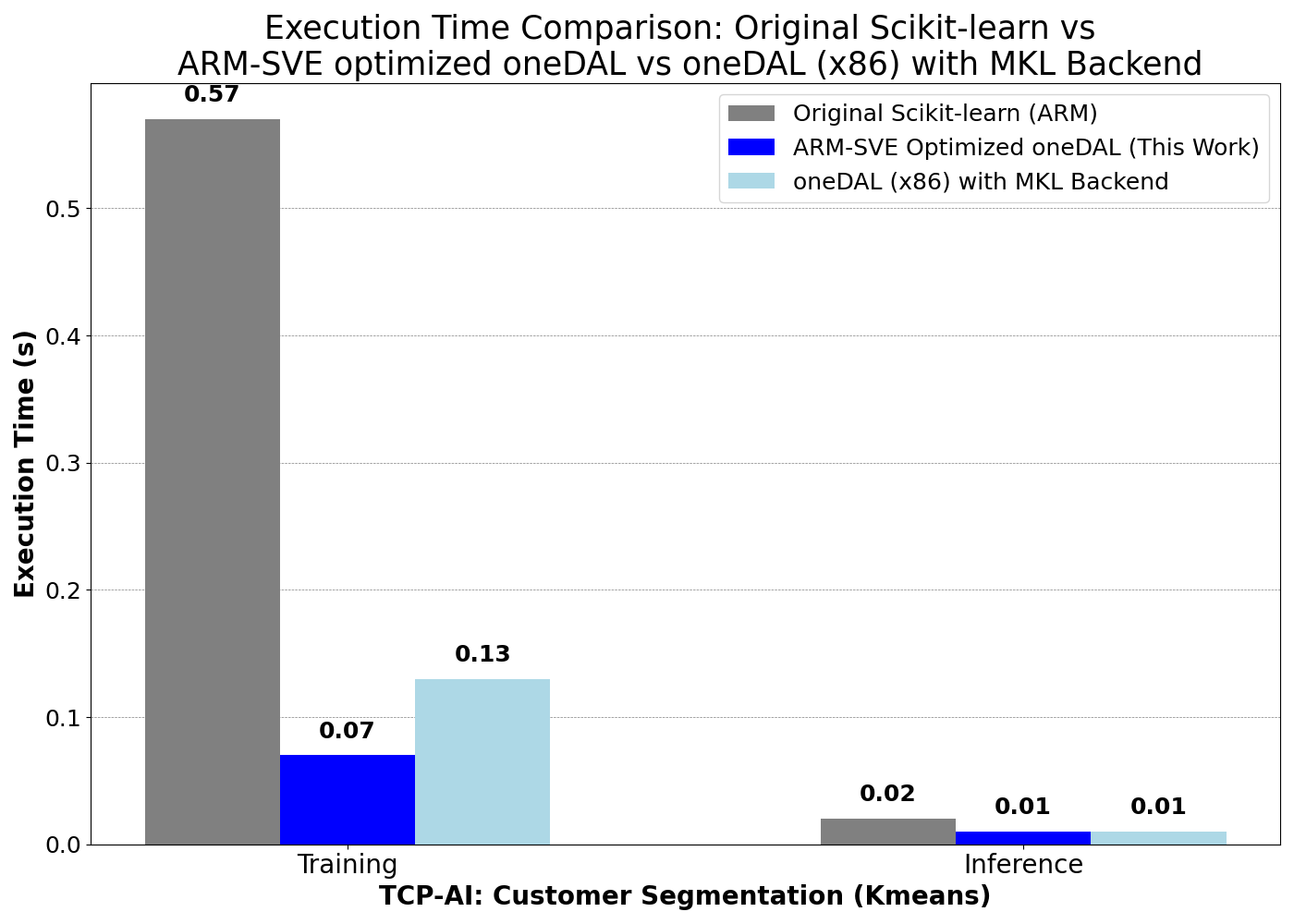}
    \caption{Performance of TPC-AI Benchmark for ARM SVE optimized \textit{oneDAL}}
    \label{fig:tpc}
\end{figure}
As shown in Figure \ref{fig:tpc}, the results for the customer segmentation task demonstrate significant performance improvements with ARM SVE optimized \textit{oneDAL} compared to the original \textit{scikit-learn} and x86 based \textit{oneDAL} with MKL backend. Our work demonstrated significant improvements during training, achieving an approximate 87.72\% reduction in execution time compared to the original \textit{scikit-learn} and 46.15\% compared to x86-based \textit{oneDAL} with MKL backend. For inference, it achieved a 50\% reduction in execution time relative to \textit{scikit-learn}, performing equal to \textit{oneDAL} on x86 with MKL. These results demonstrate that ARM-specific optimizations in \textit{oneDAL} not only significantly improve training times but also ensure competitive performance for inference, matching or surpassing x86 MKL-based implementations in both aspects.


\subsection{Real-World Use Case: Credit Card Fraud Detection}
To demonstrate the impact of our work on the ARM ecosystem, we evaluated the ARM SVE optimized \textit{ oneDAL} on a credit card fraud detection dataset consisting of 284,807 transactions, including 492 fraud cases \cite{kagledata}.

\begin{figure}[h]
    \centering
    \includegraphics[width=1\linewidth]{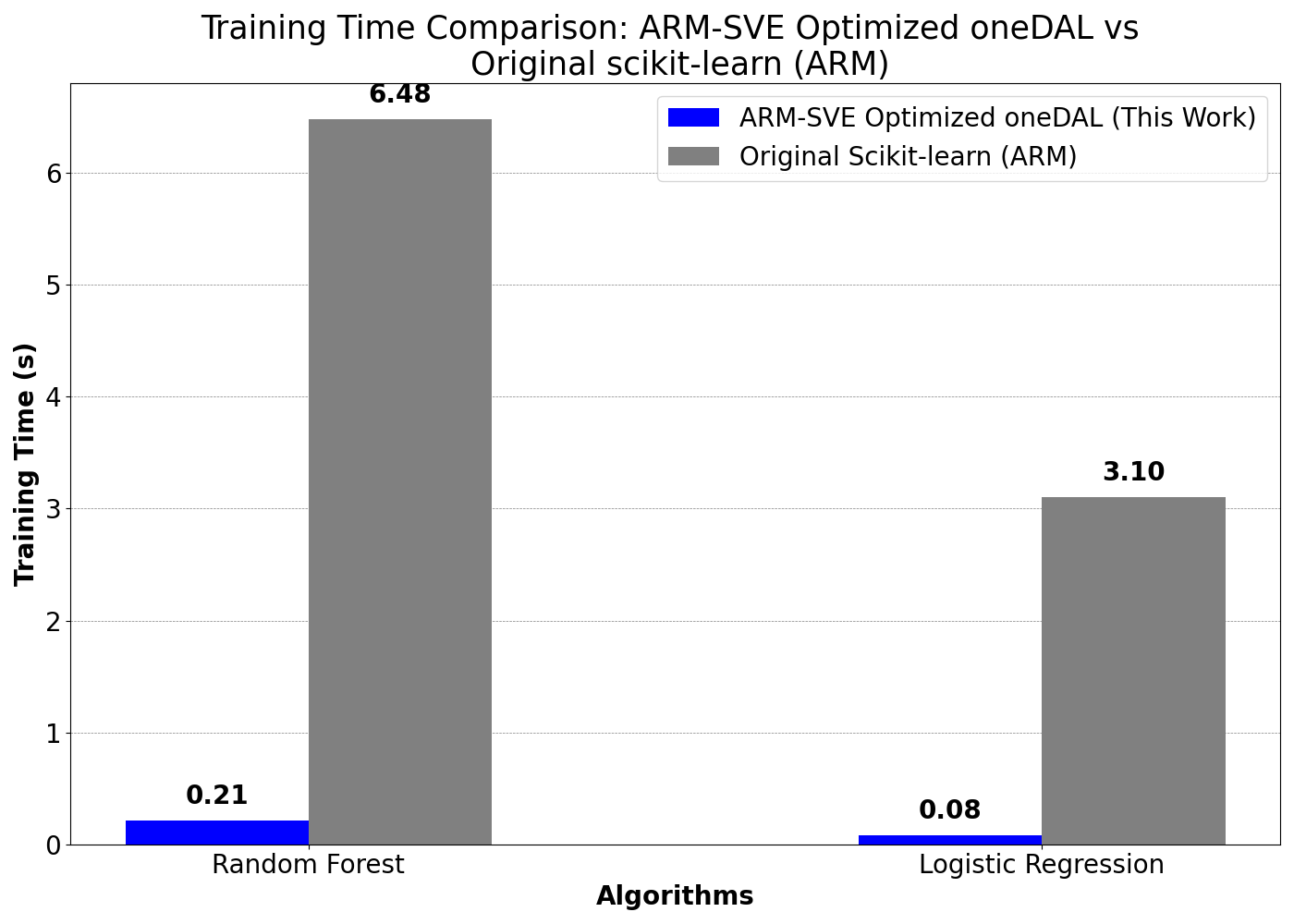}
    \caption{Performance of Credit Card Fraud Detection with ARM SVE optimized \textit{oneDAL}.}
    \label{fig:credicard}
\end{figure}

As illustrated in Figure \ref{fig:credicard}, our work achieved substantial speedups over the original \textit{scikit-learn} implementation on ARM SVE platform (Graviton3). It delivered a 31x speedup for random forest training and a 40x speedup for logistic regression compared to the original \textit{scikit-learn}. 
These results demonstrate the significant acceleration provided by our optimizations, showcasing their impact on improving ML workloads in the ARM ecosystem.

\subsection{Future Directions}

Future work could focus on optimizing oneDAL's underperforming machine learning models and also extending support to additional models to fully utilize ARM’s SVE potential in ML applications. Collaborating with open-source initiatives, such as enhancing OpenRNG, VSL, and OpenBLAS with ARM-specific optimizations, could further improve the overall performance. Additionally, developing platform-agnostic optimization techniques within \textit{oneDAL} to leverage SVE’s vector-length flexibility would ensure consistent and robust performance across diverse ARM-based systems.

\section{Conclusion}
The task of this research was to make valuable contributions to the open-source community by effectively optimizing UXL’s oneAPI oneDAL for ARM Scalable Vector Extension (SVE) architectures, using the OpenBLAS library as a reference backend. The primary goal was to achieve performance parity with the conventional x86-based oneDAL architecture, which relies on MKL, to accelerate machine learning and data analytics on ARM platforms. To achieve this, the work involved identifying dependencies, mapping functions, refactoring code, modifying the build system, implementing sparse BLAS, and developing a novel VSL to ensure compatibility with ARM architecture. Extensive ARM-specific optimizations were carried out, including vectorization analysis and the use of ARM SVE intrinsics, particularly for Support Vector Machines (SVM), and were validated through rigorous testing and validation processes. The study demonstrates that ARM architecture has the potential to provide high performance, contributing to the establishment of a more competitive and diversified HPC ecosystem.

The results demonstrate the effectiveness of the proposed approach, as validated by various benchmarks and real-world applications. The ARM-optimized \textit{oneDAL} achieved comparable or superior performance relative to existing methods. Data science algorithms, including regression, classification, and clustering, showed significant improvements on standard benchmarks. Real-world applications, such as credit card fraud detection and customer segmentation, also experienced substantial performance enhancements.

\section*{Acknowledgment}

The authors thank the oneAPI and oneDAL development teams at the UXL Foundation for their ongoing assistance and for their thorough documentation that helped us comprehend the nuances of the library. We also thank the OpenBLAS, ARM, and Intel communities for their invaluable contributions and for upholding a robust open-source, high-performance library that allowed us to perform this work advancements.

In addition, the authors are grateful for the supportive input and ideas of colleagues and collaborators who came in at each stage of the study. We also thank the HPC community for creating a collaborative atmosphere that leads to innovative thinking. Such interactions have enriched this research through learning as well as sharing thoughts with like-minded peers.

\sloppy

\end{document}